\documentclass[twocolumn,final]{jetpl}

\usepackage{graphicx}

\title{Josephson Effect in a Coulomb-blockaded SINIS Junction}
\rtitle{Josephson Effect in a Coulomb-blockaded SINIS Junction}
\sodtitle{Josephson Effect in SINIS Junction}

\author{P.\,M.\,Ostrovsky\thanks{e-mail: ostrov@itp.ac.ru} and M.\,V.\,Feigel'man}
\rauthor{P.\,M.\,Ostrovsky, M.\,V.\,Feigel'man}
\sodauthor{Ostrovsky, Feigel'man}

\address{L. D. Landau Institute for Theoretical Physics, Kosygina 2, Moscow, 119334, Russia}

\abstract{The problem of Josephson current through Coulomb-blocked nanoscale superconductor-normal-superconductor
structure with tunnel contacts is reconsidered. Two different contributions to the phase-biased supercurrent
$I(\varphi)$ are identified, which are dominant in the limits of weak and strong Coulomb interaction. Full expression
for the free energy valid at arbitrary Coulomb strength is found. The current derived from this free energy
interpolates between known results for weak and strong Coulomb interaction as phase bias changes from $0$ to $\pi$. In
the broad range of Coulomb strength the current-phase relation is substantially non-sinusoidal and qualitatively
different from the case of semi-ballistic SNS junctions. Coulomb interaction leads to appearance of a local minimum in
the current at some intermediate value of phase difference applied to the junction.}

\PACS{73.21.-b, 74.50.+r, 74.45.+c}

\begin{document}

\maketitle

\begin{figure}
\centering
\includegraphics[width=0.9\columnwidth]{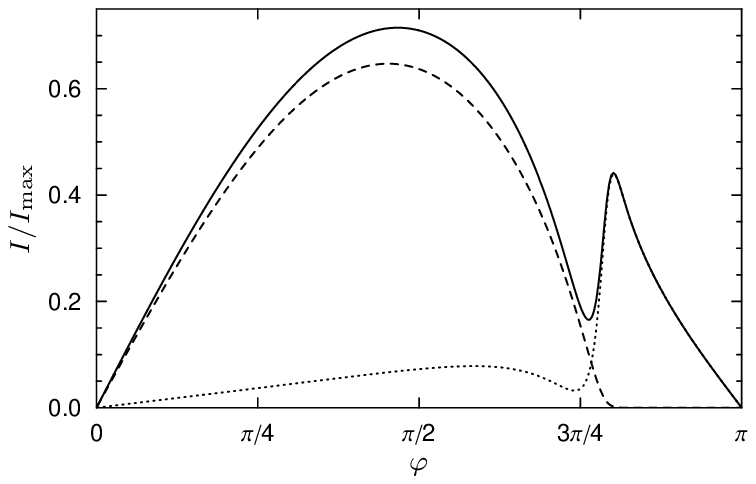}\\[12pt]
\includegraphics[width=0.9\columnwidth]{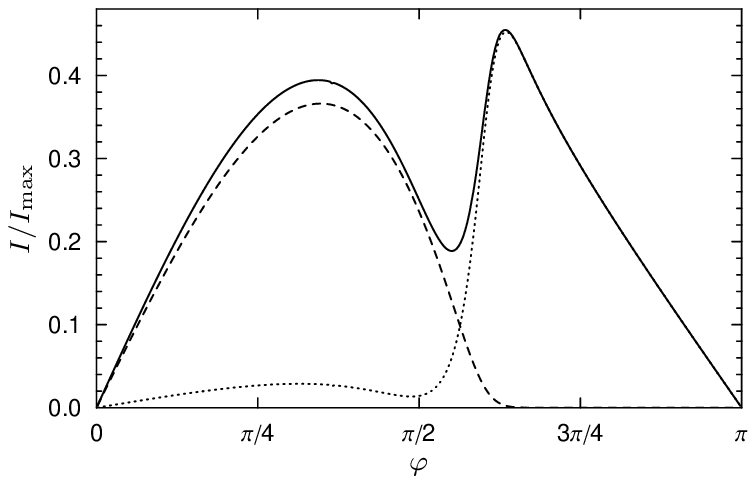}\\[12pt]
\includegraphics[width=0.9\columnwidth]{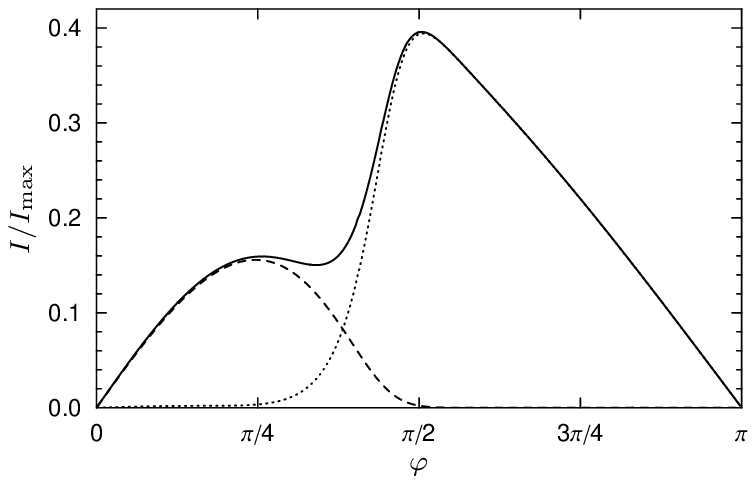}\\[12pt]
\caption{Fig 1. The dependence of Josephson current $I$ on phase bias $\varphi$. The current is normalized by its
maximal value that is reached at $\varphi = \pi/2$ in the absence of Coulomb interaction $I_{\mathrm{max}} =
(e\delta/4\hbar) G_L G_R \log(8\Delta/\delta\sqrt{G_L^2 + G_R^2})$. Dashed lines show the dependence~(\protect\ref{I0})
found in~\protect\cite{OF} without fluctuational correction. This correction is depicted by the dotted lines, while
solid curves are sums of the two contributions. The three plots correspond to different relative strength of Coulomb
blockade: (a) weak interaction $E_C\delta/E_g^2(0) = 0.5$, (b) intermediate $E_C\delta/E_g^2(0) = 1.5$, (c) strong
blockade $E_C\delta/E_g^2(0) = 2.5$. For all three plots we assume symmetric junction with $G_L = G_R = 20$ and
$\Delta/\delta = 3000$. As $\varphi$ approaches $\pi$ the Coulomb interaction always becomes strong and the current is
dominated by the fluctuational contribution~(\protect\ref{Ifl}). }
\label{Fig_current}
\end{figure}

The Josephson current in the contact of two bulk superconductors through a small normal grain (SINIS structure) was
recently studied in~\cite{OF} within the saddle-point approximation for the effective action functional which
described~\cite{OSF} superconductive proximity effect in the presence of Coulomb interaction. We found in~\cite{OF}
that Coulomb blockade in the grain results in a very special Josephson current dependence on the phase difference
$\varphi$ (see dashed lines in Fig.~\ref{Fig_current}). As $\varphi$ approaches $\pi$ the proximity effect in the grain
is suppressed, Coulomb blockade becomes relatively strong, and spectral minigap induced in the normal grain becomes
exponentially small.  This lead us to erroneous conclusion that supercurrent through the grain is exponentially weak as
well. In fact, as was pointed out to us by I.\,S.~Beloborodov and A.\,V.~Lopatin~\cite{BL}, supercurrent in the SINIS
structure may flow without any spectral minigap at all. The same result was obtained originally by C.~Bruder, R.~Fazio
and G.~Sch\"on in~\cite{BFS} by means of perturbative analysis that is valid as long as charging energy $E_C = e^2/2C$
is much larger than the proximity-induced minigap (denoted below as $\tilde{E}_g$). This additional (with respect to
our result in~\cite{OF}) contribution to the supercurrent is due to fluctuational diffuson and Cooperon modes in the N
grain, as explained below.

In~\cite{OF} the replicated dynamical sigma-model was used. The Coulomb interaction in the grain is taken into account
in the framework of the adiabatic approximation developed in~\cite{OSF}. The key point of this approximation is the
separation of energy scales: the electric potential of the grain fluctuates at frequency much larger than the proximity
induced minigap. This assumption is valid provided $E_C \gg \delta$, where $\delta$ is average level spacing in the
grain per one spin projection. The saddle point of the sigma-model gives the free energy of the system $F_0(\varphi)$;
the current is than calculated using the identity $I_0 = (2e/\hbar) \partial F_0/\partial\varphi$.

The above-mentioned additional contribution to the supercurrent is due to the fluctuations near the saddle point of the
sigma-model. Below we present a somewhat more general result calculating fluctuational correction to the total free
energy of the system. The supercurrent being the derivative of the free energy with respect to $\varphi$ acquires
significant correction in the regime when the saddle point itself gives exponentially small result. This, anyhow,
happens when the phase difference $\varphi$ comes close to $\pi$. Therefore, the results of~\cite{OF} and~\cite{BFS}
for the Josephson current are, in fact, valid in two limiting cases of weak and strong Coulomb effect correspondingly.

In order to find Josephson current in the full range of Coulomb/proximity ratio, it is necessary to supplement our
results presented in~\cite{OF} by fluctuational contribution. The saddle-point approximation used in~\cite{OF,OSF} is
justified by the inequality $\tilde E_g \gg \delta$. The correction due to fluctuations near the saddle point is
negligible in this limit. When the phase bias $\varphi$ is close to $\pi$ the parameter $\tilde E_g$ becomes
exponentially small and the above inequality is violated. It is this violation that makes fluctuational correction
important. However, due to the large value of junction's dimensionless conductance, it is sufficient to consider the
fluctuational contribution in the Gaussian approximation not going beyond the quadratic expansion of the action in soft
modes. In this paper we extend the approach of~\cite{OF,OSF} to allow for Gaussian fluctuations near the sigma-model
saddle point. The fluctuations are calculated on the background of a non-zero proximity-induced minigap. The resulting
dependence $I(\varphi)$ interpolates between that from~\cite{OF} and from~\cite{BFS} as $\varphi$ varies from $0$ to
$\pi$. Moreover, in the crossover region a local minimum of supercurrent appears (see Fig.~\ref{Fig_current}).

We start with the derivation of the fluctuational contribution to the free energy valid for arbitrary relative strength
of Coulomb blockade and proximity effect. Then the current-phase dependence is found by numerical differentiating. To
reduce unnecessary complications the temperature is put to zero. We assume the SINIS junction between two bulk
superconductors with tunnel contacts characterized by large (in units $e^2/\hbar$) normal conductances $G_L$ and $G_R$.
It is convenient to introduce the effective conductance
\begin{equation}
 G(\varphi)
  = \sqrt{G_L^2 + G_R^2 + 2G_LG_R\cos\varphi}.
 \label{G}
\end{equation}
Formally, one may treat the system as an SIN junction with one superconductive lead and normal conductance given by the
above expression~\cite{OF}. We quantify the proximity effect in the normal grain by the bare value of induced minigap
$E_g(\varphi) = G(\varphi)\delta/4$. This minigap is realized if the Coulomb interaction is absent.

The sigma-model for SINIS junction~\cite{OF,OSF} deals with the matrix field $\tilde Q^{ab}_{\varepsilon\varepsilon'}$
that bears two replica and two Matsubara energy (or, equivalently, imaginary time) indices along with particle-hole
structure in Nambu-Gor'kov space. Another field is the scalar phase $K^a_\tau$ dependent on the imaginary time and
replica number. The action has the form
\begin{multline}
 S[\tilde Q,K]
  = -\frac{\pi}{\delta}\mathop{\text{Tr}}\bigl(
      \varepsilon \hat\tau_3 \tilde Q
    \bigr) + \sum_a\int_0^{1/T} d\tau \Biggl\{
      \frac{\bigl( \dot K^a_\tau \bigr)^2}{4E_C}\\
      -\frac{\pi}{2}\, G(\varphi)\mathop{\text{tr}} \Bigl[
        \tilde Q^{aa}_{\tau\tau}\bigl(
          \hat\tau_1 \cos 2K^a_\tau + \hat\tau_2 \sin 2K^a_\tau
        \bigr)
      \Bigr]
    \Biggr\}.
 \label{action}
\end{multline}
The symbol $\hat\tau_i$ is used for Pauli matrices operating in Nambu-Gor'kov space. The operator `Tr' implies
summation over all indices including Matsubara energies and replicas, while `tr' denotes trace in Nambu-Gor'kov space
only.

The adiabatic approximation allows to integrate out the field $K$ assuming $\tilde Q$ to be fixed. The steady matrix
$\tilde{\underline Q}$ is diagonal in energies and trivial in replicas
\begin{equation}
 \tilde{\underline Q}\vphantom{Q}^{ab}_{\varepsilon\varepsilon'}
  = 2\pi \delta^{ab} \delta(\varepsilon - \varepsilon')\,
    \frac{\varepsilon\hat\tau_3 + \tilde E_g\hat\tau_1}{\sqrt{\varepsilon^2 + \tilde E_g^2}}.
\end{equation}
Here $\tilde E_g$ is the proximity-induced minigap in the normal grain renormalized by the Coulomb interaction. The
value of $\tilde E_g$ will be determined self-consistently later.

Using the adiabatic approximation we derive the effective Hamiltonian that determines the dynamics of phase $K$
\begin{gather}
 H
  = E_C\left[
      -\partial^2/\partial K^2 - 2 q \cos 2K
    \right],
 \label{ham}
 \\
 q
  = \frac{E_g(\varphi) \tilde E_g}{E_C \delta} \log \frac{2\Delta}{\tilde E_g}.
 \label{q}
\end{gather}
The parameter $q$ quantifies the relative strength of Coulomb blockade~\cite{OSF} in comparison with proximity effect.
In the limit $q \gg 1$ Coulomb interaction is effectively weak and can be treated perturbatively, wile in the opposite
case $q \ll 1$ Coulomb blockade destroys the proximity effect up to an exponentially small correction. The value of $q$
is determined self-consistently along with $\tilde E_g$.

At zero temperature $K$ is frozen in the ground state of the Hamiltonian~(\ref{ham}) with energy $E_0(q)$. Then the
total free energy acquires the form
\begin{equation}
 F_0(\varphi)
  = - \frac 1{\delta} \int \frac{\varepsilon^2\, d\varepsilon}{\sqrt{\varepsilon^2 + \tilde E_g^2}}
    + E_0(q).
 \label{freeen}
\end{equation}
The divergent integral is to be regularized by subtracting its value for the ``normal'' state with $\tilde E_g = 0$. In
all subsequent analysis we assume this regularization to be done.

The minigap $\tilde E_g$ is set by the condition $\partial F_0/\partial\tilde E_g = 0$. This gives the self-consistency
equation
\begin{equation}
 \frac{\tilde E_g}{E_g(\varphi)}
  = -\frac{1}{2 E_C}\frac{\partial E_0}{\partial q}
  = \left< 0 | \cos 2K | 0 \right>.
 \label{selfconsist}
\end{equation}
Last expression implies average value at the ground state of~(\ref{ham}). Together with~(\ref{q}) this equation forms a
closed system that determines $q$ and $\tilde E_g$.

In~\cite{OF} we calculated the supercurrent using the identity $I_0 = (2e/\hbar) \partial F_0/\partial\varphi$. The
result was
\begin{equation}
 I_0(\varphi)
  = \frac{e\delta}{4\hbar}\biggl(\frac{\tilde E_g}{E_g}\biggr)^2 G_LG_R\sin\varphi \log\frac{2\Delta}{\tilde E_g}.
 \label{I0}
\end{equation}
This dependence of supercurrent on $\varphi$ is shown in Fig.~\ref{Fig_current} by dashed lines.

To take into account fluctuations of $\tilde Q$ near the found saddle point (for detailed calculation
see~\cite{disser}) we use the parametrization of $\tilde Q$ as rotated $\hat\tau_1$ matrix: $\tilde Q = V^{-1}
e^{-iW/2} \hat\tau_1 e^{iW/2} V$. This choice is motivated by the fact that $\tilde{\underline Q} = \hat\tau_1$ at
$\varepsilon = 0$. Below we'll see that different modes of fluctuations near the saddle point, diffusons and Cooperons,
decouple in this representation. The diagonal in energies and replicas matrix $V$ is determined by the identity
$\tilde{\underline Q} = V^{-1} \hat\tau_1 V$ and expressed as $V = \cos(\pi/4 - \theta/2) - i\hat\tau_2 \sin(\pi/4 -
\theta/2)$ with $\theta$ being a standard Usadel angle, $\tan\theta = \tilde E_g/\varepsilon$, dependent on Matsubara
energy. The matrix $W$ describes deviations of $\tilde Q$ from the saddle point $\tilde{\underline Q}$. $W$
anti-commutes with $\hat\tau_1$ and hence contains two components $W^{ab}_{\varepsilon\varepsilon'} = \hat\tau_3
d^{ab}_{\varepsilon\varepsilon'} + \hat\tau_2 c^{ab}_{\varepsilon\varepsilon'}$. Below the (off-) diagonal in Nambu
space element $d^{ab}_{\varepsilon\varepsilon'}$ ($c^{ab}_{\varepsilon\varepsilon'}$) is referred to as diffuson
(Cooperon) mode. However, these modes are only analogs of standard diffuson and Cooperon that describe fluctuations
near normal metallic saddle point with no minigap.

Now we substitute the above parametrization into~(\ref{action}) and expand the action to the second order in $W$. The
result is a sum of three terms
\begin{equation}
 S[W,K]
  = S_0[K] + S_1[W,K] + S_2[W,K],
 \label{Sexpand}
\end{equation}
where $S_0[K]$ is the action corresponding to the Hamiltonian~(\ref{ham}). The terms $S_{1,2}[W,K]$ are linear and
quadratic in $W$ respectively; the explicit expressions for them can be found in~\cite{disser}. Our strategy is to
integrate out the phase $K$. As the fluctuations near the saddle point are assumed to be small, we treat the last two
terms of~(\ref{Sexpand}) as a perturbation to the bare action $S_0[K]$ or, equivalently, to the
Hamiltonian~(\ref{ham}). Thus we expand the statistical weight $e^{-S[W,K]}$ to the second order in $S_1$ and to the
first order in $S_2$. Integration with respect to $K$ implies averaging of these terms at the ground state
of~(\ref{ham}). Once the integral is calculated we rewrite the result in the form of a single exponent
\begin{gather}
 \int DW\,DK\,e^{-S[W,K]}
  = e^{-\frac{NF_0}{T}}\int DW\,e^{- S^{(1)} - S_0^{(2)} - S_{\text{int}}^{(2)}},
 \\
 S^{(1)}
  = \langle S_1 \rangle,
 \quad
 S^{(2)}_0
  = \langle S_2 \rangle,
 \quad
 S^{(2)}_{\text{int}}
  = -\frac{\langle S_1^2 \rangle - \langle S_1 \rangle^2}{2}.
\end{gather}
The integral of the term $e^{-S_0}$ yields a $W$-independent factor to the partition function corresponding to the free
energy $F_0$ calculated at the saddle point. The value of $F_0$ is given by~(\ref{freeen}) while $N$ is the number of
replicas. The symbol $\langle\ldots\rangle$ denotes the average with respect to the ground state of the
Hamiltonian~(\ref{ham}). To calculate the terms $S^{(1)}$ and $S^{(2)}_0$ we use the identity $\langle \sin 2K \rangle
= 0$ while the average value of $\cos 2K$ is determined by~(\ref{selfconsist}). The self-consistency equation provides
$S^{(1)} = 0$ as it should be at the saddle point. The two remaining terms describe fluctuations near the saddle point.
They can be written in the form
\begin{multline}
 S^{(2)}_0
  =\sum_{a,b}\int\frac{d\varepsilon\,d\varepsilon'}{8\pi\delta} \left(
      \sqrt{\varepsilon^2 + \tilde E_g^2}
      +\sqrt{\varepsilon'^2 + \tilde E_g^2}
    \right) \\
    \times \Bigl(
      c^{ab}_{\varepsilon\varepsilon'}c^{ba}_{\varepsilon'\varepsilon}
      +d^{ab}_{\varepsilon\varepsilon'}d^{ba}_{\varepsilon'\varepsilon}
    \Bigr),
 \label{S0}
\end{multline}
\begin{multline}
 S^{(2)}_{\text{int}}
  = -\frac{G^2}{2} \sum_{a} \int \frac{d\varepsilon\, d\varepsilon'\, d\omega}{(2\pi)^3} \\
    \times \Bigl[
      \lambda_c(\varepsilon,\varepsilon'; \omega)\,
      c^{aa}_{\varepsilon, \varepsilon'} c^{aa}_{\varepsilon' + \omega, \varepsilon + \omega} \\
      + \lambda_d(\varepsilon,\varepsilon'; \omega)\,
      d^{aa}_{\varepsilon, \varepsilon'} d^{aa}_{\varepsilon' + \omega, \varepsilon + \omega}
    \Bigr].
 \label{S2int}
\end{multline}
In the last expression we use the following notations
\begin{gather}
 \lambda_c(\varepsilon,\varepsilon'; \omega)
  = \frac{\pi^2}{4}\, X(\varepsilon - \varepsilon')
    \cos\tfrac{\theta_\varepsilon + \theta_{\varepsilon'}}{2}\,
    \cos\tfrac{\theta_{\varepsilon + \omega} + \theta_{\varepsilon' + \omega}}{2}, \\
 \lambda_d(\varepsilon,\varepsilon'; \omega)
  = \frac{\pi^2}{4}\, Y(\varepsilon - \varepsilon')
    \cos\tfrac{\theta_\varepsilon - \theta_{\varepsilon'}}{2}\,
    \cos\tfrac{\theta_{\varepsilon + \omega} - \theta_{\varepsilon' + \omega}}{2}, \\
 X(\omega)
  = \sum_{n > 0} \bigl|\bigl<0\bigl|\cos 2K\bigr|n\bigr>\bigr|^2
    \frac{2(E_n - E_0)}{\omega^2 + (E_n - E_0)^2}, \label{X} \\
 Y(\omega)
  = \sum_{n > 0} \bigl|\bigl<0\bigl|\sin 2K\bigr|n\bigr>\bigr|^2
    \frac{2(E_n - E_0)}{\omega^2 + (E_n - E_0)^2}. \label{Y}
\end{gather}
The two functions $X(\omega)$ and $Y(\omega)$ appeared from the averaging of $S_1^2$. They are nothing but the Fourier
components of irreducible correlators $\langle\!\langle \cos 2K_0 \cos 2K_\tau \rangle\!\rangle$ and $\langle\!\langle
\sin 2K_0 \sin 2K_\tau \rangle\!\rangle$ respectively. In the expressions~(\ref{X}) and~(\ref{Y}) we use
$\left|i\right>$ and $E_i$ to denote eigenvectors and eigenvalues of~(\ref{ham}).

With all these definitions in hand we employ the standard replica trick to calculate the free energy
\begin{equation}
 F
  = F_0 + T\lim_{N \to 0} \frac{1}{N} \left[
    1 - \int DW\, e^{- S_0^{(2)} - S_{\text{int}}^{(2)}}
  \right].
 \label{Ftotal}
\end{equation}
The last term of this expression is the fluctuational contribution. It contains Gaussian integral with rather
complicated quadratic form in the exponent. The term $S_0^{(2)}$ is fully diagonal with respect to both replica and
Matsubara energy indices of $c$ and $d$ components. The complication arises from the $S_{\text{int}}^{(2)}$ term where
different energies are coupled~(\ref{S2int}). The Gaussian integration is equivalent to computing the determinant of
this quadratic form. To find the value of this determinant we use the standard trick~\cite{AGD}. Let us consider the
derivative of free energy with respect to $G^2$: $\partial F/\partial(G^2)$. The factor $G^2$ is present explicitly
only in the $S_{\text{int}}^{(2)}$ term. Then the derivative is
\begin{equation}
 \frac{\partial F}{\partial(G^2)}
  = \lim_{N \to 0} \frac{T}{NG^2} \int DW\, S_{\text{int}}^{(2)}\, e^{- S_0^{(2)} - S_{\text{int}}^{(2)}}.
 \label{FG2int}
\end{equation}
The pre-exponent of the integrand is quadratic in $c$ and $d$. Hence the differentiating with respect to $G^2$ reduced
the problem of calculating the determinant to the calculation of the inverse matrix elements. They correspond to the
two-particle Green functions, that are Cooperon and diffuson. The Cooperon is
\begin{multline}
 \bigl<
   c^{ab}_{\varepsilon, \varepsilon'}
   c^{pq}_{\varepsilon' + \omega', \varepsilon + \omega}
 \bigr>
  = \raisebox{-9pt}{\begin{picture}(86,20)(0,0)
      \put(17,18){\line(1,0){40}}
      \put(17,6){\line(1,0){40}}
      \multiput(21,18)(2,0){15}{\line(1,-3){4}}
      \put(15,4){\llap{$\scriptstyle \varepsilon,\, a$}}
      \put(15,17){\llap{$\scriptstyle \varepsilon'\!\!,\, b$}}
      \put(59,4){$\scriptstyle \varepsilon + \omega,\, q$}
      \put(59,17){$\scriptstyle \varepsilon' + \omega'\!\!,\, p$}
    \end{picture}} \\
  = 2\pi \delta^{aq} \delta^{bp} \delta(\omega - \omega')\,
    C^{ab}(\varepsilon,\varepsilon'; \omega).
 \label{Cooperon}
\end{multline}
The analogous diffuson function is similar to the above average with $c$ components being replaced by $d$. We denote
this propagator by $D^{ab}(\varepsilon,\varepsilon'; \omega)$. Angle brackets in the last expression imply the average
with the Gibbs weight given by the quadratic action $S_0^{(2)} + S_{\text{int}}^{(2)}$. This averaging also contains
normalization by the partition function that is the determinant we are calculating. However, this normalization factor
is canceled when the $N \to 0$ limit is taken in~(\ref{FG2int}). (Note, the replica trick was originally invented
namely for this cancelation.)

Below we concentrate on the quantity $C^{ab}(\varepsilon,\varepsilon'; \omega)$. Another propagator is found in
analogous way; the only change is the replacement of $\lambda_c$ by $\lambda_d$. To find
$C^{ab}(\varepsilon,\varepsilon'; \omega)$ we have to solve a simple Dyson equation
\begin{equation}
 \raisebox{-3pt}{\begin{picture}(36,12)
   \put(3,12){\line(1,0){30}}
   \put(3,0){\line(1,0){30}}
   \multiput(7,12)(2,0){10}{\line(1,-3){4}}
 \end{picture}}
  = \raisebox{-3pt}{\begin{picture}(36,12)
      \put(3,12){\line(1,0){30}}
      \put(3,0){\line(1,0){30}}
    \end{picture}}
    +
    \raisebox{-3pt}{\begin{picture}(78,12)
      \put(3,12){\line(1,0){30}}
      \put(3,0){\line(1,0){30}}
      \put(33,0){\line(1,1){12}}
      \put(33,12){\line(1,-1){12}}
      \put(39,6){\circle*{4}}
      \put(45,12){\line(1,0){30}}
      \put(45,0){\line(1,0){30}}
      \multiput(49,12)(2,0){10}{\line(1,-3){4}}
    \end{picture}},
 \label{Dyson}
\end{equation}
where the bare correlator $C_0(\varepsilon,\varepsilon')$ is determined by inverse eigenvalues of the diagonal
quadratic form $S_0^{(2)}$ and the vertex is the matrix element of $S_{\text{int}}^{(2)}$
\begin{gather}
 C_0(\varepsilon,\varepsilon')
  = \raisebox{-3pt}{\begin{picture}(33,16)
      \put(3,12){\line(1,0){27}}
      \put(3,0){\line(1,0){27}}
      \put(12,1){$\scriptstyle \varepsilon,\, a$}
      \put(12,13){$\scriptstyle \varepsilon'\!\!,\, b$}
    \end{picture}}
  = \frac{\delta}{\pi}\,
    \frac{1}{\sqrt{\varepsilon^2 + \tilde E_g^2} + \sqrt{\varepsilon'^2 + \tilde E_g^2}},
 \\
 \raisebox{-9pt}{\begin{picture}(57,23)
   \put(17,6){\line(1,1){12}}
   \put(17,18){\line(1,-1){12}}
   \put(23,12){\circle*{4}}
   \put(15,4){\llap{$\scriptstyle \varepsilon,\, a$}}
   \put(15,17){\llap{$\scriptstyle \varepsilon'\!\!,\, b$}}
   \put(31,4){$\scriptstyle \varepsilon + \omega,\, a$}
   \put(31,17){$\scriptstyle \varepsilon' + \omega,\, b$}
 \end{picture}}
  = G^2 \delta^{ab} \lambda_c(\varepsilon,\varepsilon'; \omega).
\end{gather}
The solution to the equation~(\ref{Dyson}) is
\begin{multline}
 C^{ab}(\varepsilon,\varepsilon'; \omega)
  = C_0(\varepsilon,\varepsilon') \biggl[
      2\pi\delta(\omega) \\
      + \delta^{ab} \frac{G^2 \lambda_c(\varepsilon,\varepsilon'; \omega)}
                    {1 - G^2 {\mathcal C}(\varepsilon - \varepsilon')}\,
        C_0(\varepsilon' + \omega, \varepsilon + \omega)
    \biggr].
 \label{C}
\end{multline}
In the denominator of the last term the screening function ${\mathcal C}(2 \Omega)$ appears
\begin{multline}
 {\mathcal C}(2\Omega)
  = \int \frac{d\varepsilon}{2\pi}\,
    \lambda_c(\varepsilon + \Omega,\varepsilon - \Omega; 0)
    C_0(\varepsilon + \Omega,\varepsilon - \Omega) \\
  = \frac{\delta}{8}\, X(2\Omega) \left[
      \log\frac{2\Delta}{\tilde E_g}
      - \sqrt{1 + \frac{\tilde E_g^2}{\Omega^2}}\,
        \mathop{\mathrm{arcsinh}} \frac{\Omega}{\tilde E_g}
    \right].
 \label{sC}
\end{multline}
This integral contains logarithmically divergent contribution that gives the first term in square brackets. The second
term comes from small values $\varepsilon \lesssim \max\{\Omega, \tilde E_g\}$. The analogous screening function for
the correlator $D^{ab}(\varepsilon,\varepsilon'; \omega)$ is
\begin{equation}
 {\mathcal D}(2\Omega)
  = \frac{\delta}{8}\, Y(2\Omega)\left[
      \log\frac{2\Delta}{\tilde E_g}
      - \frac{\mathop{\mathrm{arcsinh}} (\Omega / \tilde E_g)}{\sqrt{1 + \tilde E_g^2 / \Omega^2}}
    \right].
 \label{sD}
\end{equation}

Now everything is ready for the calculation of the derivative $\partial F/\partial(G^2)$. The pre-exponent
in~(\ref{FG2int}) contains the sum over single replica index. The saddle point is trivial in replica space. Therefore,
this sum will be canceled by $1/N$. Another feature of pre-exponent is the factor $2\pi\delta(0)$. This factor appears
from the delta-function in the definition of Cooperon~(\ref{Cooperon}). The same is also true for the diffuson term in
the pre-exponent. At finite temperature $2\pi\delta(0) = 1/T$ that cancels temperature in~(\ref{FG2int}). Once this
cancelation is established we can safely assume $T = 0$.

After substitution of~(\ref{C}) and similar expression for $D^{ab}(\varepsilon,\varepsilon'; \omega)$
into~(\ref{FG2int}) and integration with respect to $\omega$ and to the sum $\varepsilon + \varepsilon'$ we come to a
single integral over $\Omega = \varepsilon - \varepsilon'$
\begin{equation}
 \frac{\partial F}{\partial(G^2)}
  = -\int \frac{d\Omega}{4\pi}\, \left[
      \frac{{\mathcal C}(\Omega)}{1 - G^2 {\mathcal C}(\Omega)}
      + \frac{{\mathcal D}(\Omega)}{1 - G^2 {\mathcal D}(\Omega)}
    \right].
 \label{FG2}
\end{equation}
In the limit $G^2 = 0$ the vertex part of the action, $S_{\text{int}}^{(2)}$, is absent. This leads to the absence of
fluctuational contribution at $G^2 = 0$ in the limit $N \to 0$. Using this fact we finally come to the full expression
for the free energy by integrating~(\ref{FG2})
\begin{equation}
 F
  = F_0 + \int \frac{d\Omega}{4\pi}\, \log
    \bigl[ 1 - G^2 {\mathcal C}(\Omega) \bigr]
    \bigl[ 1 - G^2 {\mathcal D}(\Omega) \bigr].
 \label{Fcomplet}
\end{equation}

The total Josephson current is now easy to find by differentiating free energy $I = (2e/\hbar) \partial
F/\partial\varphi$. The current derived from the first term $F_0$ was found in~\cite{OF}. Rather simple
expression~(\ref{I0}) exists for this quantity. The fluctuational contribution is much more complicated. The dependence
on phase difference $\varphi$ is contained not only in the factor $G^2$ according to~(\ref{G}) but also in the
screening functions. The situation is very much simpler in the physically interesting limit of strong Coulomb blockade.
The minigap is strongly suppressed and the saddle point itself produces negligible contribution to Josephson current.
In the fluctuational part of the free energy we put $\tilde E_g = 0$. The expression~(\ref{ham}) becomes a free
particle Hamiltonian as $q = 0$. Only first term is left in both sums~(\ref{X}) and~(\ref{Y}), hence $X(\omega) =
Y(\omega) = 4E_C/(16E_C^2 + \omega^2)$. The screening functions~(\ref{sC}) and~(\ref{sD}) become identical and contain
only $\log(\Delta/\Omega)$ in square brackets. The dependence of free energy on $\varphi$ is now provided only by the
factor $G^2$ in~(\ref{Fcomplet}). For the Josephson current in strong Coulomb blockade regime we have
\begin{equation}
 I(\varphi)
  = \frac{e\delta}{4\hbar}\, G_L G_R\sin\varphi \log\frac{\Delta}{2E_C}.
 \label{Ifl}
\end{equation}
Thus the result of~\cite{BFS} is reproduced.

In the opposite limit of weak Coulomb blockade the fluctuational contribution to the supercurrent is small in
comparison with $I_0$. When the phase difference changes from $0$ to $\pi$ the system goes from weak to strong Coulomb
blockade regime. This means that the result~(\ref{I0}) gradually transforms into~(\ref{Ifl}). Numerical differentiating
of~(\ref{Fcomplet}) gives the solid curves plotted in Fig.~\ref{Fig_current} for the current-phase dependence. Note the
local minimum that appears in the crossover region. There is no simple analytic theory for this effect. However, this
is likely to be the most prominent feature of the system.

In conclusion, we have calculated the fluctuational correction to the free energy and to the Josephson current in
Coulomb blockaded SINIS junction. This correction plays major role in the limit of strong Coulomb interaction. At
intermediate value of phase bias a well-defined local minimum of Josephson current appears. We are grateful to
I.\,S.~Beloborodov and A.\,V.~Lopatin for pointing out the fluctuational effect and for fruitful discussions of the
calculation. This work was supported by the program ``Quantum Macrophysics'' of RAS, by the Russian Ministry of
Education and Science under the contract RI-112/001/417, and by RFBR grant 04-02-16348-a. P.M.O acknowledges the
support of ``Dynasty Foundation''.

\end{document}